\documentclass[aps,prb,citeautoscript,preprint]{revtex4}
\usepackage{amsmath}
\usepackage{amssymb}
\usepackage{amsfonts}
\usepackage{graphicx}
\usepackage{mathrsfs}
\usepackage{bm}
\usepackage{color}
\usepackage{cancel}
\usepackage{mathbbol}
\usepackage{epsfig}
\usepackage{units}
\usepackage{esint}
\usepackage{graphicx}

\newcommand{\s}[1]{\left<#1\right>}

\newcommand{\rk}[1]{\left( #1\right)}

\newcommand{\rem}[1]{}
\newcommand{\tn}[1]{\textnormal{#1}}

\begin{document}

\title{Experimental control of transport resonances in  a  coherent quantum rocking ratchet}

\author{Christopher Grossert\*}
\email{grossert@iap.uni-bonn.de}
\affiliation{Institut f\"ur Angewandte Physik der Universit\"at Bonn,
 Wegelerstr. 8, 53115 Bonn}

\author{Martin Leder}
\affiliation{Institut f\"ur Angewandte Physik der Universit\"at Bonn,
Wegelerstr. 8, 53115 Bonn}

\author{Sergey Denisov}

\affiliation{Department of Applied Mathematics, Lobachevsky State University of Nizhny Novgorod, Russia}
\affiliation{Institut f\"ur Physik, Universit\"at Augsburg,
Universit\"atsstra\ss e 1, 86159 Augsburg, Germany}
\affiliation{Sumy State University, Rimsky-Korsakov Street 2, 40007 Sumy, Ukraine}

%\affiliation{Sumy State University, Rimsky-Korsakov Street 2,
%40007 Sumy, Ukraine}
\author{Peter H\"anggi}
\affiliation{Institut f\"ur Physik, Universit\"at Augsburg,
Universit\"atsstra\ss e 1, 86159 Augsburg, Germany}
\affiliation{Department of Applied Mathematics, Lobachevsky State University of Nizhny Novgorod, Russia}
\affiliation{Nanosystems Initiative Munich, Schellingstr. 4,
D-80799 M\"{u}nchen, Germany}
%\affiliation{Center for Phononics and Thermal Energy Science and
%School of Physical Science and Engineering, Tongji University,
%200092 Shanghai, Peoples Republic of China}
\author{Martin Weitz}
\affiliation{Institut f\"ur Angewandte Physik der Universit\"at Bonn,
Wegelerstr. 8, 53115 Bonn}

\maketitle

\textbf{The ratchet phenomenon is a means to get directed transport without
net forces. Originally conceived  to rectify stochastic motion
and describe operational principles of biological motors,
the ratchet effect can be used to  achieve controllable coherent quantum transport.
This transport is an important ingredient of several perspective quantum devices including
atomic chips. Here we examine coherent transport of ultra-cold atoms in a rocking quantum ratchet.
This is realized by loading a rubidium atomic
Bose-Einstein condensate into a periodic  optical  potential subjected
to a biharmonic temporal drive. The achieved
long-time coherence  allows us to resolve resonance enhancement of the
atom transport induced by avoided crossings in the Floquet spectrum of the system.
By tuning the strength of the temporal
modulations, we observe a bifurcation of a single resonance into a
doublet. Our measurements reveal the important role
of interactions among Floquet eigenstates for quantum ratchet transport.
}
\newline

A controllable dissipationless, fully coherent quantum transport of ultra-cold atoms is
a prerequisite for several applications, ranging from quantum
information processing with atom chips \cite{chips1,chips2} to
high-precision BEC-gravimetry \cite{inter1,inter2}. There are several ways
to reach this goal \cite{means1,means2,means3} and the ratchet effect is one of
them \cite{ketzmerick1,ketzmerick2, den,quantum_ratchet,pr}. The essence of
this effect is that a particle in a periodic potential can be set
into a  directed motion by using
zero-mean time-periodic modulations of the potential only
\cite{rpr,rmp2009}.

There exists a variety of different ratchet devices
\cite{rpr,rmp2009}, with setup-sensitive conditions for occurrence of
directed transport. Of prime importance in this context is the
identification of the dynamical symmetries  which \textit{prevent}
the appearance of the directed motion \cite{pr}. A proper choice
of the  system parameters, especially of the driving field,
leads to the breaking of all no-go symmetries to yield  an average net
current.

There are two popular Hamiltonian ratchet setups for both, classical \cite{rpr,rmp2009} and
quantum systems \cite{Gong2006,summy1,summy2,Wang2008,sadgrove1,sadgrove2,hoogerland,pr}.
While flashing ratchets are characterized
by multiplicative driven potentials, $U(x) = V(x) F(t)$, rocking ratchets are realized with
periodically tilted potentials, $U(x) = V(x) +  F(t)x$. In the flashing mode of operation the
forcing enters multiplicative, whereas it is of additive character for the rocking mode. As a consequence,
the two setups belong to different dynamical symmetry groups \cite{pr}. Particularly, the rocking ratchet can be realized with a single-harmonic potential while a flashing ratchet needs a
potential with at least two spatial harmonics \cite{pr}.

The symmetry analysis alone, however, fails to predict the transport
direction and its average velocity. These quantities
depend on the inherent  mechanisms specific to
the system's nature and  control parameters. Physical intuition
may sometimes apply, for example, a high velocity can be expected
in the case of resonant driving, when the modulating frequency
matches the characteristic frequency of the potential, as was
verified with  experiments using cold-atoms in the regime of
classical ratchets \cite{renzoni1,renzoni2,renzoni3} and, as well, with a
flashing quantum ratchet realized with a Bose-Einstein condensate of rubidium atoms \cite{quantum_ratchet}.

An intriguing  phenomenon was  predicted in
numerical simulations of quantum coherent  ratchets
\cite{den}. Namely, the ratchet current can be substantially
boosted by tuning specific Floquet states of a periodically driven
potential into an avoided crossing \cite{wigner,fac,holt}. It was
also predicted that these transport resonances follow an universal
bifurcation scenario upon increasing the driving strength. The
scenario is dictated by generic properties of the Floquet spectra
of quantum ratchets. This theoretical result provides a
possibility of a more subtle (as compared with the symmetry-based
scheme) control of the quantum ratchet transport.
The Floquet resonances were theoretically observed with both above mentioned driving schemes
\cite{den}. However, an experimental verification  requires  a regime of coherent quantum transport
on time scales much larger then the period of the driving.

Our objective here is the resolution of the theoretically-predicted
Floquet resonances in experiment, by using an ac-driven optical potential
and an atomic Bose-Einstein condensate. In contrast to the previous experiment using a quantum flashing ratchet \cite{quantum_ratchet},
where a biharmonic potential building upon the dispersion of multi-photon Raman transitions was
used, the rocking setup requires only  a standard sinusoidal
standing-wave optical potential. For alkali atoms with a
$s$-electronic ground state configuration $L=0$, the absence of
the second harmonic in the optical potential is beneficial because it allows for much
longer coherence times as compared to those achieved  with the
flashing setup. Therefore, by implementing the rocking scheme,
we can observe Floquet resonances in the mean
velocity of ultra-cold atoms and the splitting of a single resonance into a
doublet of transport resonances.

\section*{Results}

\subsection*{\textit{Experimental realization}}
The quantum rocking ratchet is described by the time-periodic
Hamiltonian \cite{den,pr}
\begin{equation}
\hat{H} = \hat{p}^{2}/2m + V_0\cos(2k\hat{x}) - F(t)\hat{x},
\label{Eq:ham}
\end{equation}
where $m$ denotes the mass of the atom, $k=2\pi/\lambda$ is the wave-vector of
the potential, where $\lambda \simeq \unit[783.5]{nm}$ is the wavelength of the
laser beams used in the experiment, and $V_0$ is the tunable lattice depth.
A time-periodic force,
$F\rk{t}=F\rk{t+T}$, is implemented by modulating one of the two
counter-propagating lattice beams with a time-dependent frequency
$\Omega\rk{t}$.
In the lab frame, this field produces a moving lattice potential,
$V\rk{x',t}=V_0\cos[2kx'- f\rk{t}]$, $f\rk{t}=\int_{t_0}^{t+t_0} \Omega\rk{s}ds$,
where the temporal evolution starts at the (starting) time $t_0 \in [0, T]$. This parameter specifies the strength of the
rocking force when the modulations are switched on.
In the co-moving frame, this corresponds to a stationary potential   subjected to a
rocking inertial force $F(t) = \frac{m}{2k}\ddot f\rk{t}$, see (\ref{Eq:ham}). Similar to the setup in
Refs.~\cite{renzoni1,renzoni2,renzoni3} we employ a biharmonic frequency modulation
\begin{equation}
\Omega\rk{t}=\Omega_0\{\sin[\omega_{\tn{m}}t]+\beta\sin[2
\omega_{\tn{m}}t+\theta]\}, \nonumber
\end{equation}
where $\Omega_0$ denotes the modulation amplitude, $\omega_{\tn{m}}$ is the
modulation frequency, and $\beta$ and $\theta$ are the relative amplitude and relative phase of the
second harmonic. Thus, in the co-moving frame this corresponds to the  Hamiltonian in (\ref{Eq:ham})
with a rocking force $F(t)$  of the form \cite{den}
\begin{eqnarray}\nonumber
F\rk{t}&=&A_1\cos[\omega_{\tn{m}}t] + A_2\cos[2\omega_{\tn{m}} t+\theta],
\label{Eq:force}
\end{eqnarray}
with $A_1=\frac{m}{2k}\Omega_0\omega_{\tn{m}}$ and
$A_2=\frac{m}{k}\Omega_0\beta\omega_{\tn{m}}$.

In our experiment, a Bose-Einstein condensate (BEC) of $^{87}$Rb atoms is
produced first in the $m_F=0$ spin projection of the $F=1$ hyperfine
ground state by evaporative cooling. After that  the condensate
expands freely for $3${ms} and converts the internal
interaction energy of the dense atomic cloud
into kinetic energy. From the resulting velocity distribution, a narrow slice
of the momentum width $\Delta p$ = $0.2 \hbar k$ is separated with a
$330\mu$s long Raman pulse, transferring atoms into the $m_F = -1$ spin
projection state. The atoms are then loaded into a rocked periodic potential
formed by an optical standing wave, detuned $3$nm to the red end of the rubidium
$D2$-line.
After the interaction with the optical potential, the atomic cloud is allowed to expand
freely for $\unit[15-20]{ms}$ and then an absorption image is recorded. By using time-of-flight (TOF)
images, we analyze the velocity distribution of the atomic
cloud, see Fig. 1. The interaction with the lattice potential during a time span $[t_0, t_0 + t]$ results in a
diffraction pattern with a set of discrete  peaks, separated by two
photon recoils, with the $n$th order peak corresponding to a momentum of
$p_n = 2\hbar k n$; see Fig. 1 (a). The mean momentum of the atomic cloud is calculated
as $\bar{p}= \sum_n b_n  p_n$, where $b_n$ is the relative
population of the $n$th momentum state. Due to finiteness of the contrast and
sensitivity of the imaging system, we restrict the summation to $n=-3, \ldots, 3$.

\subsection*{\textit{Dependence of atom current on the modulation starting time}}
The atomic current produced by the rocking quantum ratchet, Eqs. (\ref{Eq:ham}-\ref{Eq:force}), can be evaluated
in terms of the eigenfunctions of the operator which propagates the system over one period
of the driving, $U(T)$,  $|\psi_{\alpha}(T)\rangle=\exp(-i \varepsilon_{\alpha}T/\hbar)$
$|\phi_{\alpha}(T)\rangle$. These eigenfunctions are stroboscopic snapshots of
the time-periodic Floquet states $\{ |\phi_{\alpha}(t) \}_{\alpha = 1,2,...}$,
$|\phi_{\alpha}(t+T)\rangle= |\phi_{\alpha}(t)\rangle$ \cite{shirley,sambe,Grifoni}.
In the lab  frame, the system Hamiltonian is a spatially-periodic operator
and its reciprocal space is spanned by the quasienergy Bloch bands,
$\varepsilon_{\alpha}(\kappa)$, $\kappa \in
[-\pi/L, \pi/L]$,  $\alpha = 1, 2, ...$, and the Floquet states are
parameterized by the quasimomentum values $\kappa$, $|\phi_{\alpha,\kappa}(t)\rangle$ \cite{kolovsky}.
A non-vanishing transport is expected for $\kappa =  0$ when $A_1,A_2 \not = 0$ and
$\theta\not=l\cdot\pi,l\in\mathbb{Z}$.   This choice of parameters
results in the breaking of the sole dynamical symmetry,
\begin{eqnarray}
{\hat S}_t: \{x, \hat{p}, t; t_0\} \rightarrow \{x, -\hat{p}, -t;-t_0\},  \label{quasi}
\end{eqnarray}
preventing the de-symmetrization of the  eigenstates
\cite{ketzmerick1, ketzmerick2, den,quantum_ratchet}. The average velocity of $\alpha$-th
Floquet state $|\phi_{\alpha, \kappa}(t)\rangle$ at quasimomentum $\kappa$ is
determined by the local slope, $\upsilon_{\alpha,\kappa} =
\hbar^{-1}\partial\varepsilon_{\alpha} (\kappa) / \partial\kappa$
\cite{ketzmerick1, ketzmerick2, den}.
An initial wave packet can be expanded over the instantaneous Floquet basis,
$|\psi(t_0)\rangle = \int_{-\infty}^{\infty} \big[ f(\kappa)   \sum_{\alpha} C_{\alpha,\kappa}
(t_0) |\phi_{\alpha, \kappa}(t_0)\rangle \big]  d\kappa$. The distribution $f(\kappa)$ is determined by the
momentum profile of the initial wave packet, which is transformed into the profile in the $\kappa$-space.
The velocity after an overall interaction
time $t$ then reads \cite{pr}
\begin{eqnarray}
%\nonumber
v(t;t_0) &=& \int_{-\infty}^{\infty} \big[ f(\kappa)   \sum_{\alpha}|
C_{\alpha,\kappa}(t_0)|^2 \upsilon_{\alpha,\kappa} \big] d\kappa + v_{\rm beat}(t;t_0) = v_{\rm a}(t_0) + v_{\rm
beat}(t;t_0),\label{Eq:current}
\end{eqnarray}
where the last term on the rhs accounts for the interference between different
Floquet states. Its time average disappears in the asymptotic limit,
$\lim_{t\rightarrow\infty} \,\langle v_{\rm beat}(t;t_0) \rangle_{t} \rightarrow 0$, provided that either (i)
there are at least several Floquet states which overlap substantially with the initial wave-function
that is well-localized at $\kappa = 0$, $f(\kappa) \approx \delta(\kappa)$ \cite{den}, or (ii)
the initial wave packet is spread over the quasimomentum space (we discuss the corresponding
mechanism in the next section). In the latter case it is enough to have
two Floquet bands effectively overlapping with the initial wave packet; this latter situation is the case in our experiment;
see Fig. 2 (c). The theoretical quantity $m \cdot v(t,t_0)$ should be compared with the mean momentum
of the atomic cloud, $\bar{p}_{\theta}(t;t_0)$, a quantity measured in the experiment
and defined in the previous section.

Because of the explicit time-dependence of the Floquet states, the weights
$C_{\alpha,\kappa}(t_0)$ depend on the starting time $t_0$, so that the asymptotic
velocity depends on the starting time even when the initial wave-function
and all other parameters are held fixed \cite{den,quantum_ratchet}. We first studied
this quantum feature, namely the dependence of the atomic transport on the starting
time $t_0 \in [0, T]$, $T = 2\pi/\omega_{\tn{m}}$. Our Fig. 1 (b) depicts the
experimental results for two values of
$\theta$, $\pi/18$ (filled blue dots) and $17\pi/18$
(filled green squares). In both cases we observe a strong dependence of the
ratchet transport on the starting time $t_0$. Theory predicts a particular
symmetry, reading, $\bar p_{\pi-\theta}(t;T/2-t_0)= \bar p_{\theta}(t;t_0)$.
This symmetry follows from the invariance of the quantum Hamiltonian (\ref{Eq:ham} - \ref{Eq:force})
under the transformation of $\theta$ and $t_0$, combined with the double reversal
$\{t,x\} \rightarrow \{-t,-x\}$ and complex conjugation. The result of this transformation
applied to the experimentally measured momentum dependencies is depicted with
Fig. 1 (c). Within experimental uncertainty, the momentum dependencies
perfectly match each other. We interpret this finding
as  key evidence for the coherent character of the dynamics of our quantum ratchet.

\subsection*{\textit{Temporal evolution of the mean atomic momentum}}
Figure 2 (a) depicts the mean atomic momentum
$\s{\bar p}$, where $\s{...}$ denotes the averaging of the
momentum over the starting time $t_0$, versus the number of
modulation periods. The interference between the contributing Floquet
states comes into a play immediately after the switch-on of the modulations
and induces the appearance of a non-zero current already after
several periods of the modulations. Upon increasing elapsing interaction time $t$, the mean momentum
exhibits several oscillations, as expected from the interference beating (note the
last term on the rhs of Eq.~(\ref{Eq:current}), and saturates towards a nearly constant value. The initial
wavefunction of the loaded BEC can be effectively represented as a coherent
superposition of two Floquet states, see Fig. 2 (c). The spectral gap
at the avoided crossing point (near $\kappa = 0$), $\omega_{\rm beat} = \delta\varepsilon/\hbar$,
$\delta\varepsilon = |\varepsilon_{\alpha} - \varepsilon_{\beta}|$, specifies the time scale of
the interference beating. The theoretical model yields $\omega_{\rm beat} = 0.04\, \omega_{\tn m}$, cf.  Fig. 2 (c),
which matches the time, $t_{\rm max} \approx 20T$, after which the first maximum appears
in the dependence $\langle \bar{p}(t)\rangle$ versus interaction time  $t$.

The finite momentum dispersion of the BEC produces an additional, while fully
coherent, damping-like effect for the time evolution of the
current. Namely, contributions of Floquet eigenstates from different
$\kappa$ - bands, characterized by continuously  changing
quasienergies, $\varepsilon_{\alpha}(\kappa)$, cf.  Fig. 2 (c), result upon elapsing interaction time in a
self-averaging of the interference term $v_{\rm beat}$ towards
zero. Thus, the finite momentum width of the initial packet
removes the need for the additional run-time averaging of the
current in order to obtain the asymptotic velocity $v_{\rm a}(t_0)$ [this averaging was used in Ref. \cite{den}
when calculating ratchet dynamics of the wave packet $f(\kappa) = \delta(\kappa)$].
For a broader momentum BEC slice, this effect causes with increasingly elapsing interaction time $t$ a substantial damping of
the oscillations, note the filled green squares in  panel Fig. 2 (a).

\subsection*{\textit{Detection of transport resonances}}
We next turn to the  issue of quantum transport resonances present in coherently rocking
quantum ratchets, with both harmonic amplitudes $A_1,A_2 \not= 0$. A theoretical analysis is elucidative in the
limit $f(\kappa) = \delta (\kappa)$ (the analysis for the general case
can be performed by using the recipe in Ref.~\cite{fei} ).
When the phase $\theta = l\cdot\pi,~l\in\mathbb{Z}$, the system (\ref{Eq:ham}-\ref{Eq:force}) obeys
time-reversal symmetry so that all  Floquet states become non-transporting,
$\upsilon_{\alpha,\kappa=0} \equiv 0$. An asymptotic current is absent though a
transient current is still possible due to the above-discussed interference effects. From the symmetry
analysis of the Schr\"{o}dinger equation with the Hamiltonian (\ref{Eq:ham}-\ref{Eq:force}),
it follows that the dependence of the averaged (over $t_0$)
asymptotic velocity $v_{\rm a}(\theta) = \langle  v_{\rm a}(\theta;t_0) \rangle$ on $\theta$
obeys $v_{\rm a}(\theta) = -v_{\rm a}(\theta \pm \pi) = -v_{\rm a}(-\theta)$ \cite{den}.
The results of experimental measurements nicely fit this theoretical prediction, see Fig. 2 (b).

Theoretically one may  find \cite{den} a resonant-like  increase
of the average current versus $\theta$ when tuning the amplitude of the driving.
All Floquet states are ordered with respect to their
averaged kinetic energy in  ascending order $\alpha = 1,...$.
The Floquet state $\phi_{1}(t)$ has the lowest kinetic energy and
any initial wave function which has a lower kinetic energy overlaps  with this state mainly.
This state is strongly affected by the change of the potential shape and
thus the corresponding dependence $\varepsilon_{1}(\theta)$ exhibits noticeable
dispersion upon the variation of $\theta$, with a 'tip', either minimum-like or maximum-like, at the point
of maximal asymmetry, $\theta = \pm \pi/2$; note the the bottom sketch in Fig. 3 (a).
%This follows from the symmetry analysis which predict that the quasienergy $\theta$-bands posses the property
%\begin{eqnarray}
%\varepsilon_{\alpha}(\kappa=0, \theta) = \varepsilon_{\alpha}(\kappa=0, -\theta) =
%\varepsilon_{\alpha}(\kappa=0,\theta \pm \pi) \,.
%\label{Eq:quasi_sym}
%\end{eqnarray}
%The term `$\theta$-band' denotes here the continuous line produced by
%the quasienergy value of a particular Floquet state upon a variation
%of the parameter $\theta$ in the system Hamiltonian.
Floquet states with high kinetic energies  possess large average velocities $\bar{\upsilon}$.
We call them 'ballistic states'. The quasienergy dependence on $\theta$ of a typical ballistic state,
$\varepsilon_{\alpha=n}(\theta)$, $n \gg 1$, is close to a  straight line because the state is only weakly affected by
the variations of the potential shape.

Even when being distant on the energy scale,
the two bands, $\alpha = 1$ and $\alpha = n$, can be brought into an avoided crossing on the quasienergy scale,
$\varepsilon_{\alpha} \in [-\hbar \omega/2,\hbar \omega/2]$,
by tuning the amplitudes of the modulating force, $A_1$ and $A_2$ \cite{holt, den}.
Due to the parabolic-like structure of the dependence $\varepsilon_{1}(\theta)$,
the two states always meet first (if they do) at the points $\theta = \pm \pi/2$, see second
(from the bottom) sketch in  Fig. 3 (a). The eigenstates mix at these
avoided crossings \cite{reichl}, so that their wave functions exchange
their structures. This effect leads to an increase of the average velocity of the Floquet
state with minimal kinetic energy. Because the crossing is forbidden, a further increase of the
modulation strength causes a bifurcation of the avoided crossing point into two
avoided crossing points, with the latter moving apart upon even further increase, as it shown on third and
forth (from the bottom) sketches on Fig. 3 (a). For an initial wave function which substantially overlaps  with
the Floquet state assuming minimal kinetic energy the  'mixing' of the Floquet states will reveal
itself through a resonance-like behavior of the velocity dependence $\langle v_{\rm a}\rangle$ \cite{den}.
The particular choice of the initial low-energy wavefunction, for example the zero-plane wave $|0\rangle$ or the ground
state of the stationary potential $V(x)$, is not essential because it does modify the results only
slightly. The avoided crossing should not be sharp, however, otherwise the beating
time $t_{\rm beat} = 2 \pi\hbar/\triangle\varepsilon$ will be larger than the timescale of the experiment
and the mixing effect cannot be detected.

Fig. 3 (b)  depicts the mean amomentum of the atomic
cloud as a function of the phase $\theta$ for different values of the amplitude
$\Omega_0$. For  small values of $\Omega_0$, we observe
an enhancement of transport at the point $\theta = \pi/2$. It is attributed to a local
contact of Floquet states as indicated on the two lower panels on
Fig. 3 (a).  For larger values of the  amplitude $\Omega_0$, the peak
splits into a doublet. This bifurcation  is attributed to the splitting
of a single avoided crossing point as depicted on the top panel of
Fig. 3 (a).

We also performed numerical simulations of the ratchet dynamics by using the model Hamiltonian in
Eqs. (\ref{Eq:ham}-\ref{Eq:force}).
Figs. 4 (a, b) depicts the Floquet band structure in
$\kappa$-space at $\theta = \pi/2$ obtained for the  parameter set used in
the experiment. The width and color of the band indicate the relative
populations (additionally averaged over starting times $t_0$) of the bands for an
initial  Gaussian wave packet. The obtained numerical results confirm that for the chosen
set of parameters and the chosen initial condition the system dynamics indeed is governed
by two bands, a Floquet band exhibiting minimal kinetic energy (the corresponding band assumes a flat line)
and a ballistic band.

For the  modulation amplitude $\Omega_0=\unit[70]{kHz}$,
Fig. 4 (a), there occurs an avoided crossing between the Floquet state with minimal kinetic energy
and the ballistic state. This avoided crossing is responsible for
the appearance of the resonant single-peak  in the atomic current. On the other
hand, for $\Omega_0=\unit[122]{kHz}$ , Fig. 4 (b), there is almost
no interaction between the ballistic state and the low-energy state. This explains the
minimum in the momentum dependence at $\theta = \pi/2$, cf. the top panel on
 Fig. 3 (a). The bifurcated avoided crossings are shifted from the point
$\theta = \pi/2$; the latter  causes the formation of a double-peak pattern in
the momentum dependence $\langle \bar{p} \rangle$ versus $\theta$.

\section*{Discussion}
In conclusion, we demonstrate the control of coherent quantum
transport in a rocking quantum ratchet by engineering  avoided
crossings between Floquet states.
Rocking quantum ratchets allow for an experimentally
long-lasting coherent transport regime and thus make possible the
observation of  specific bifurcation scenarios, such as those transport resonances. Our results
give direct experimental evidence for the interaction between
Floquet states of the driven system to
determine the directed atomic current, enabling a fine-tuned
control of transport of ultra-cold matter in the fully coherent
limit.

Other problems for which
coherent controllable interactions between Floquet states are
beneficial include quantum systems containing a leak \cite{leak1,leak2,leak3}
or photonic systems with losses \cite{wiersig, leak4},
where tunable external modulations can create long-lived dynamical
modes. Periodically-modulated optical potentials can also be put
to work as tunable quantum 'metamaterials'. This scenario also allows the sculpturing
of materials with Dirac cones in the quasienergy spectrum by subtle
engineering of avoided crossings between designated Floquet states.
To achieve such Dirac
points, the corresponding avoided crossing has to be sufficiently sharp, which means the
difference between the quasienergies of the participating Floquet states,
$\Delta\varepsilon$, must assume values
smaller than the characteristic time $\tau \propto 1/v$, where $v$
is the velocity of the atomic beam moving across the optical potential.
The avoided crossing in $\kappa$-space would then be ignored during
the corresponding motion along the Floquet band \cite{holt}.
Because this condition (to have a sharp avoided crossing) is  opposite
to what we utilized in this work
(to have a broad avoided crossing, in order to resolve it on the time scale of the experiment)
this latter perspective is even more appealing. The coherence time of the order $100T$
can be sufficient to meet the above condition.
Possible applications of this idea include the study of Klein tunneling
\cite{klein1,Salger}, or also the observation of interacting
relativistic wave equations phenomena, such as a chiral confinement
\cite{merkl} in ac-driven systems.

\section*{Methods}

\subsection*{\textit{Numerical simulation}}

In order to reproduce the experimental measurements, we accounted for a finite
width of the initial wave function and performed simulations for a Gaussian
initial wave packet. If the initial packet is not too broad and well localized
within the first Brillouin zone the neglect of its tale contributions to the
overall current produces a uniform rescaling of the ratchet current. This
resolves  the issue of the finite contrast resolution when calculating
diffraction peak populations obtained in the experiments. However, it also leads
to an overestimation of the current. In order to fit the measurements, we did perform
numerical simulations with a Gaussian wave packet with a dispersion (being the fitting
parameter used in our case)
five times smaller than that used in the experiments. The region
$\kappa \in [-2\hbar k, 2\hbar k]$ was sliced with $500$ equidistant quasimomentum subspaces,
 assuming the initial state being in the form of the zero-order plane wave.
We have also performed simulations assuming the Bloch groundsate
of the undriven potential as the initial wave function (within each
$\kappa$-slice). The obtained results only slightly differ from the presented ones.
We propagate the wave functions independently and after an interaction time
$t$ sum the velocities by weighting them with the
Gaussian distribution. Similar to the experiment, these results were averaged
over eight different values of $t_0$. The  obtained dependencies are  in very
good agreement with the experimental data, see thin colored lines in Fig. 2 (a),
and in  Fig. 3 (b).

\subsection*{\textit{Experimental details}}

Our experiment uses an all-optical approach to produce a quantum degenerate
sample of rubidium atoms, which subsequently
is loaded onto a modulated optical lattice
potential to realize a rocking ratchet setup. Bose-Einstein condensation of
rubidium ($^{87}$Rb) atoms is reached by evaporative cooling of atoms
in a quasistatic optical dipole trap, formed by a tightly focused horizontally
oriented optical beam derived from a CO{$_2$}-laser with optical power
of $\unit[36]{W}$ operating near $\unit[10.6]{\mu m}$ wavelength.
A spin-polarized Bose-Einstein condensate is realized by applying an
additional magnetic field gradient
during the final stage of the evaporation, in this case a
condensate of $5\cdot 10^4$ atoms in
the $m_{\textnormal{F}}=0$ spin projection of the $F=1$ hyperfine electronic
ground state component \cite{all-optical,leder}. A homogeneous magnetic bias of
$\unit[2.9]{G}$ (corresponding to a
$\Delta\omega_{\tn{z}} \approx 2\pi \cdot \unit[2]{MHz}$ splitting between adjacent
Zeeman sublevels) is applied, which removes the degeneracy of magnetic
sublevels.

By letting the condensate expand freely for a period of $\unit[3]{ms}$,
the atomic interaction energy is converted into kinetic energy.
The measured momentum width of the condensate the atoms then reach is
$\Delta p = 0.8 \hbar k$. We subsequently use a $\unit[330]{\mu s}$
long Raman pulse to cut out a narrow slice of
$\Delta p=0.2 \hbar k$ width from the initial velocity distribution, transferring
the corresponding atoms into the the $m_F=-1$ spin projection.
The atoms are now loaded into a modulated optical lattice potential
formed by two counter-propagating optical lattice beams deriving
from a high power diode laser with output power of $\approx \unit[1]{W}$
detuned $\unit[3]{nm}$ to the red of the rubidium $D2$-line. Before irradiating
the atomic cloud, the two optical lattice beams each pass an acousto-optic
modulator and are
spatially filtered with optical fibers. One of the modulators is used to in a
phase-stable way modulate the relative frequency of the two lattice beams
with a biharmonic function. The acousto-optic
modulators are
driven with two phase locked arbitrary function generators. The maximum relative
frequency modulation amplitude ($\approx \unit[700]{kHz}$)
of the lattice beams is clearly
below the Zeeman splitting between adjacent Zeeman sublevels
($\omega_{\tn{z}} / 2\pi\approx \unit[2]{MHz}$), which suppresses unwanted Raman
transitions between the sublevels.\\
After the interaction with the driven lattice, we let the atomic cloud
expand freely for a $\unit[15-20]{ms}$ long period and subsequently measure the
population in the $F=1$,$m_{F}=-1$ state with an absorption imaging technique.
For this, the corresponding atoms are first transferred to the $F=2$, $m_{F}=-1$
groundstate sublevel with a $\unit[34]{\mu s}$ long microwave $\pi$-pulse,
and then a shadow image is recorded
with a resonant laser beam tuned to the $F=2 \rightarrow F'=3$ component
of the rubidium $D2$-line onto a CCD-camera, see also \cite{leder}.
The used time-of-flight technique allows us to analyze the velocity distribution
of the atomic cloud.

%%%%%%%%%%%%%%%%%%%%%%%%%%%%%%%%%%%%%%%%%%%%%

%%%%%%%%%%%%%%%%%%%%%%%%%%%%%%%%%%%%%%%%%%%%%%%

\section*{Acknowledgements}
This work was supported by the DFG Grants No.
We1748/20 (M.W.), HA1517/31-2 (P.H.) and DE1889/1-1 (S.D.).
The theoretical analysis was supported
by the Russian Science Foundation (Grant No. 15-12-20029) (P.H.).

\section*{Competing financial interests}
The authors declare that they have no competing financial interests.

\section*{Author contributions}
C. G. and M. L. conducted the experiment and analyzed the data. S. D. did the numerical
studies and simulations. S. D., P. H. and M. W. planned the project.
All authors contributed to the writing of the paper and to the
interpretation of the results.

\newpage
\section*{Figure captions}

\begin{figure}[htb]
\begin{center}
\includegraphics[width=0.99\textwidth]{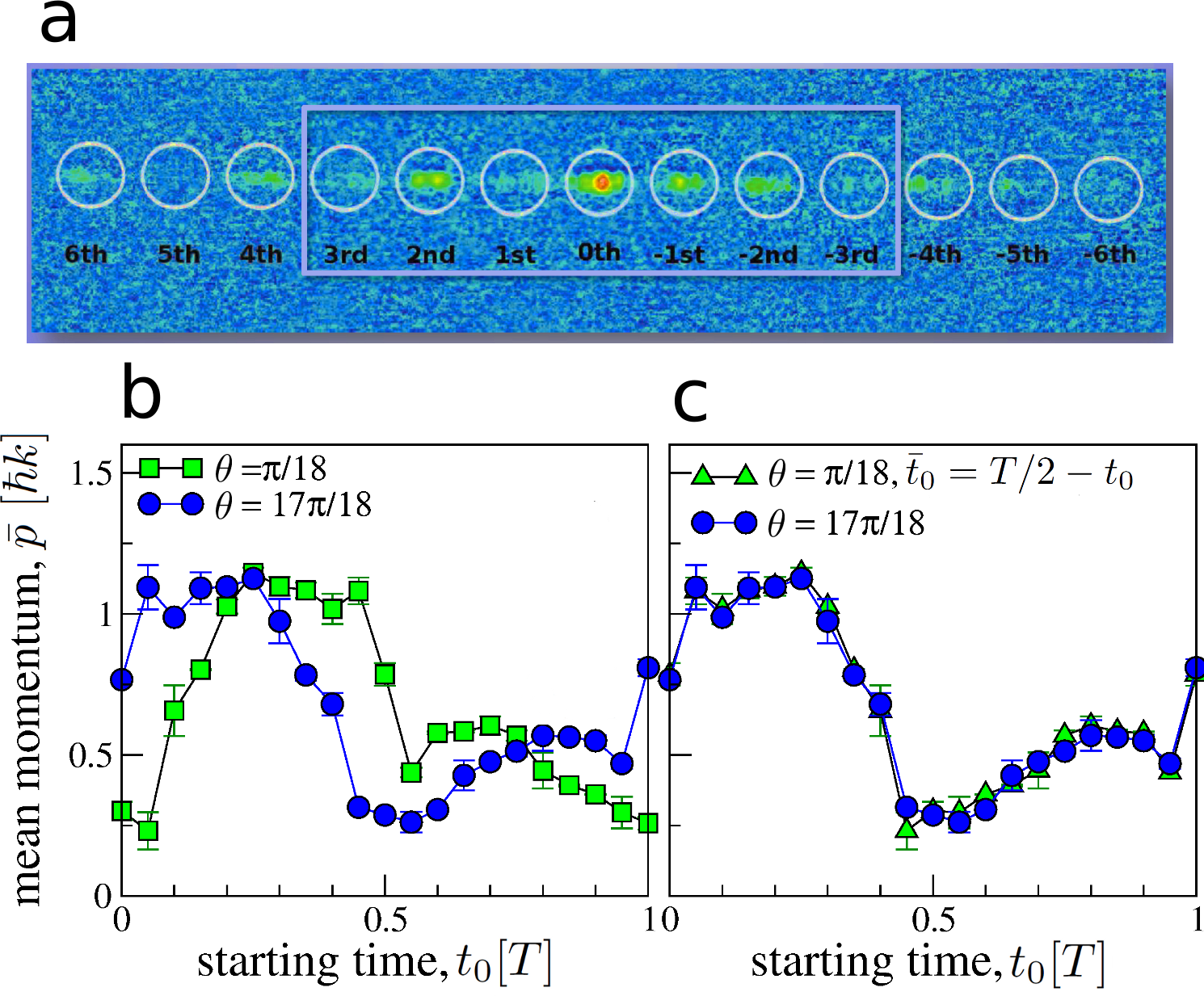}
%\caption{Dependence of computational efficiency measures $R_{100}$ (blue) and $P_{100}$ (red) on the dimension of the problem. Multi-thread version on a single cluster node ($16$ cores with shared memory) is employed (see Table~\ref{table:2}). Here $N_{\rm steps} = 10^2$,  $M = 50$.}
\label{figure:1}
\end{center}
\end{figure}

\textbf{Figure 1 | Dependence of atom transport on starting time}\\
(a): Time-of-flight image recorded after 15 ms of free expansion time, showing the atomic velocity distribution after $100$ modulation periods.
The white circles mark the position of the visible diffraction peaks.
%The population of the $n$-th momentum state
%was calculated by integrating the intensity of a peak inside the corresponding circle.
%The mean momentum was then calculated  as a sum of the `state's velocity $\times$ state's population`,over the momentum states. The summation was restricted to $n = -3,..,3$.

Lower panels: Mean atomic momentum as a function of the starting
time $t_0$ measured for two different values of $\theta$. The shown error bars correspond to the standard deviation of the mean over three measurements per data point. (b): Original
experimental data. (c): The result of the transformation $ t_0 \rightarrow (T/2- t_0) \bmod T$
applied to the data set for $\theta = 17\pi/18$.
The measurements were performed after an
interaction time of $t=100T$. The experimental parameters are $V_0 = 4.5 E_{\tn{r}}$,
$\Omega_0 = \unit[241.8]{kHz} $, $\omega_{\tn{m}} = \unit[24]{kHz}\approx 6.42\omega_{\tn{r}}$,
$\beta = 12/13\approx 0.923$.
\newpage

\begin{figure}[htb]
\begin{center}
\includegraphics[width=0.99\textwidth]{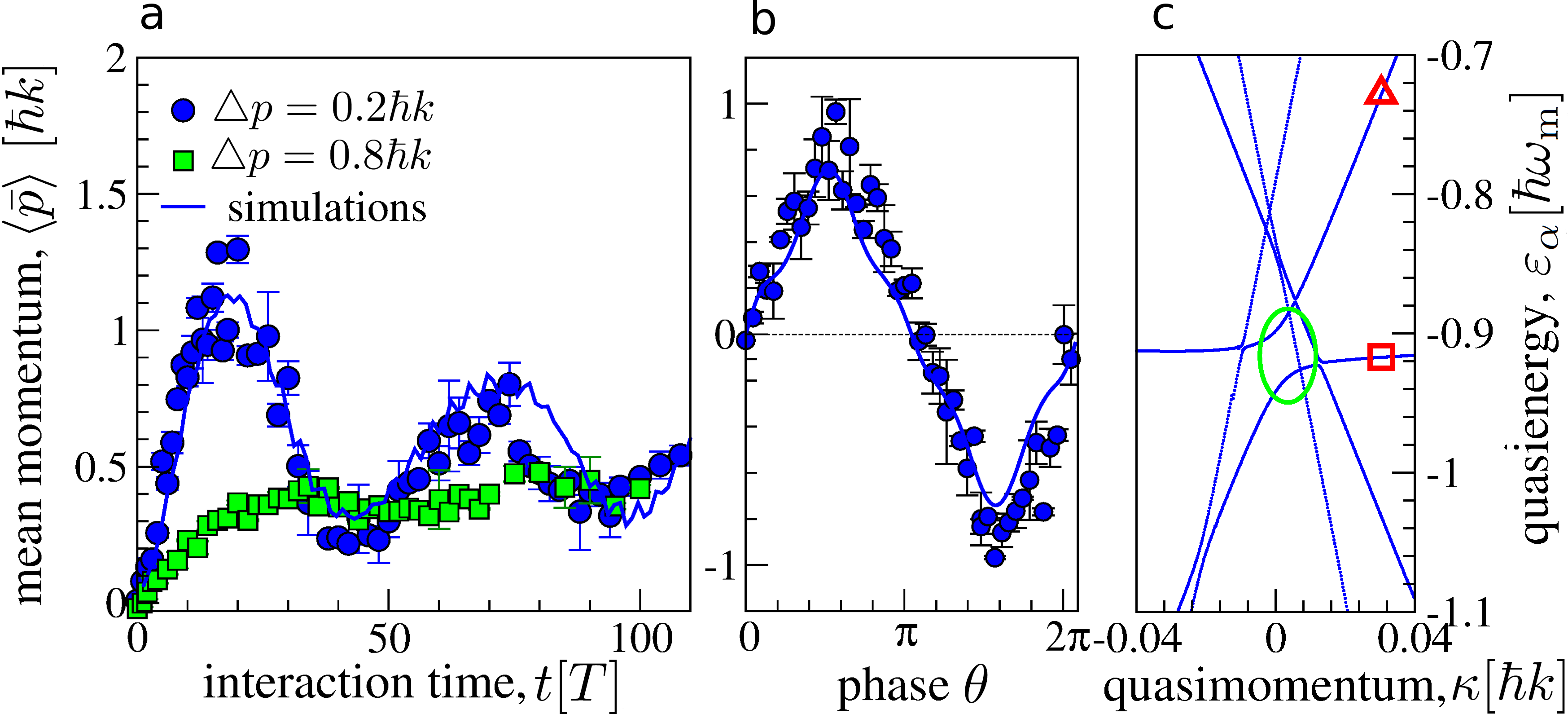}
%\caption{Dependence of computational efficiency measures $R_{100}$ (blue) and $P_{100}$ (red) on the dimension of the problem. Multi-thread version on a single cluster node ($16$ cores with shared memory) is employed (see Table~\ref{table:2}). Here $N_{\rm steps} = 10^2$,  $M = 50$.}
\label{figure:1}
\end{center}
\end{figure}

\textbf{Figure 2 | Temporal evolution of the mean atomic momentum}\\
(a): Mean atomic momentum as a function of the
interaction time $t$ for two values of the momentum dispersion $\Delta p$ of the
initial BEC slice. For every time instant the momentum was averaged over
eight equidistant values of $t_0 \in [0,T]$.
(b): Mean atomic momentum measured after an interaction time $t=70T$
as a function of relative phase $\theta$ for $\Delta p = 0.2 \hbar k$.
The data nicely obey the symmetry property in Eq. (\ref{quasi}).
The thin (blue) lines in (a) and (b) correspond to the results of numerical simulations.
The shown error bars correspond to the standard deviation of the averaged mean momentum.
The parameters used are $V_0 = 4.45 E_r$,
$\Omega_0 = \unit[93.6]{kHz}$, $\omega_{\tn{m}} = \unit[59.4]{kHz}\approx 15.9\omega_{\tn{r}}$
$\beta = 12/13\approx 0.923$ and $\theta = \pi/2$.
(c) The part of the quasienergy spectrum
near the center of the first Brillouin zone (only five bands are shown). The Floquet resonance
is produced by the avoiding crossing (marked by yellow ellipse) between two bands,
the Floquet band with the minimal kinetic energy ($\Box$)
and a ballistic band ($\triangle$). These corresponding two Floquet states dominantly contribute  to
the velocity, Eq.~(\ref{Eq:current}).
\newpage

\begin{figure}[htb]
\begin{center}
\includegraphics[width=0.99\textwidth]{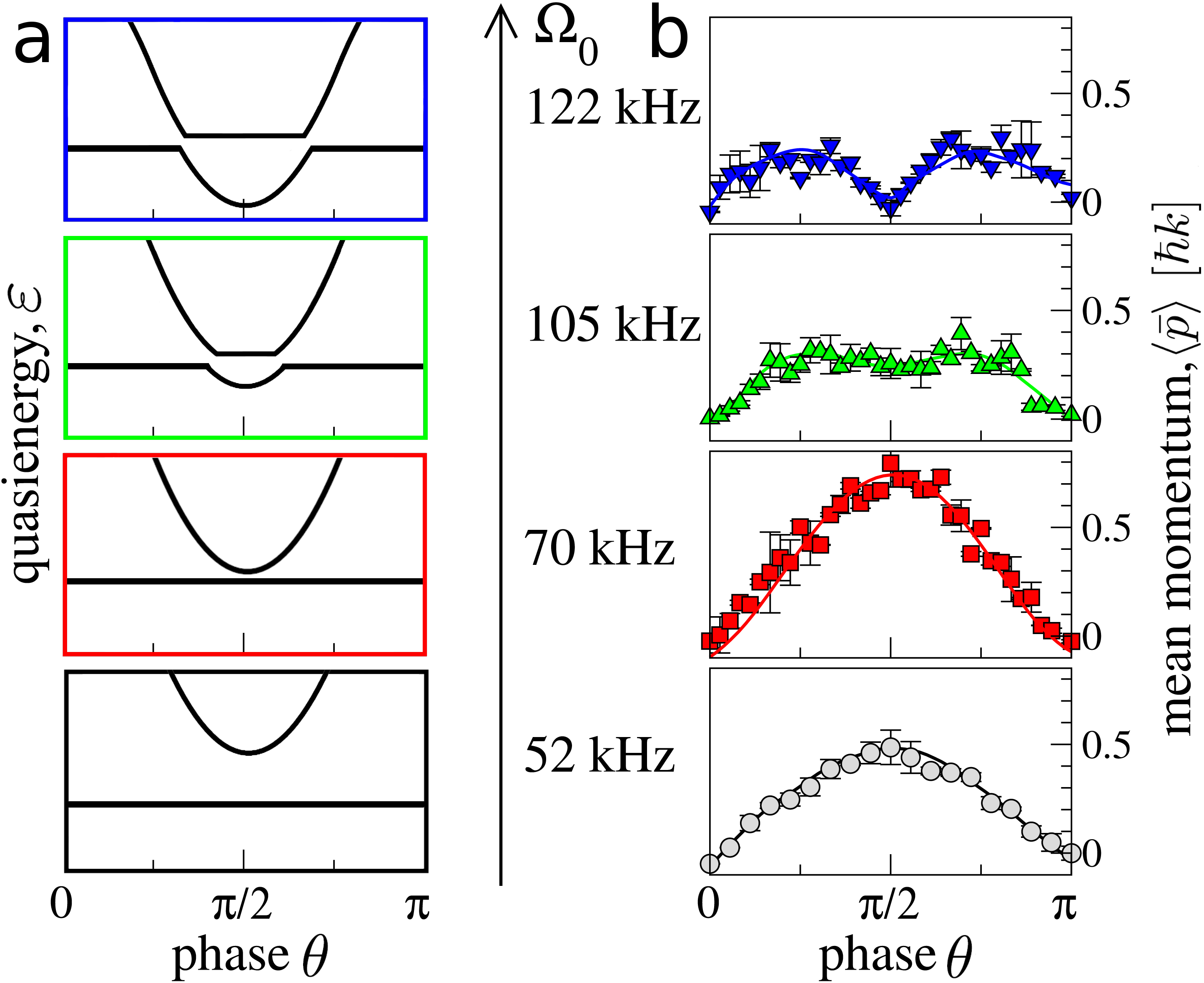}
%\caption{Dependence of computational efficiency measures $R_{100}$ (blue) and $P_{100}$ (red) on the dimension of the problem. Multi-thread version on a single cluster node ($16$ cores with shared memory) is employed (see Table~\ref{table:2}). Here $N_{\rm steps} = 10^2$,  $M = 50$.}
\label{figure:1}
\end{center}
\end{figure}

\textbf{Figure 3 | Bifurcation of a transport resonance}\\
(a): A sketch of the interaction scenario between
 quasienergy Floquet bands (bottom to top). For low values of the modulation
amplitude, the Floquet ground band (upper parabolic curve) lies far from a ballistic
band (straight line). Upon increasing the modulation amplitude, the tip of
the Floquet ground band approaches the ballistic band and touches the latter at the
points of the maximal asymmetry $\theta = \pi/2$. Because the crossing is forbidden, a further tuning
of the parameter causes a bifurcation of the avoided crossing point
into  two avoided crossing points (for the sake of clarity,
the smallness of the avoided crossings  are exaggerated). The color
of the frames corresponds to the coloring used for the right panel.
(b): Mean atomic momentum as a function of $\theta$ for different values
of the modulation amplitude $\Omega_0$. The measurements were performed after the
interaction time $t = 70T$ and averaged over eight equidistant values of
$t_0 \in [0,T]$. The thin lines correspond to numerical results obtained for
the Gaussian initial wave-packet. The shown error bars correspond to the standard
deviation of the averaged mean momentum. The experimental parameters are
$V_0 = 3.55 E_{\tn{r}}$, $\beta = 13/7\approx1.86$ and $\omega_{\tn{m}}=16\omega_{\tn{r}}$.
\newpage

\begin{figure}[htb]
\begin{center}
\includegraphics[width=0.99\textwidth]{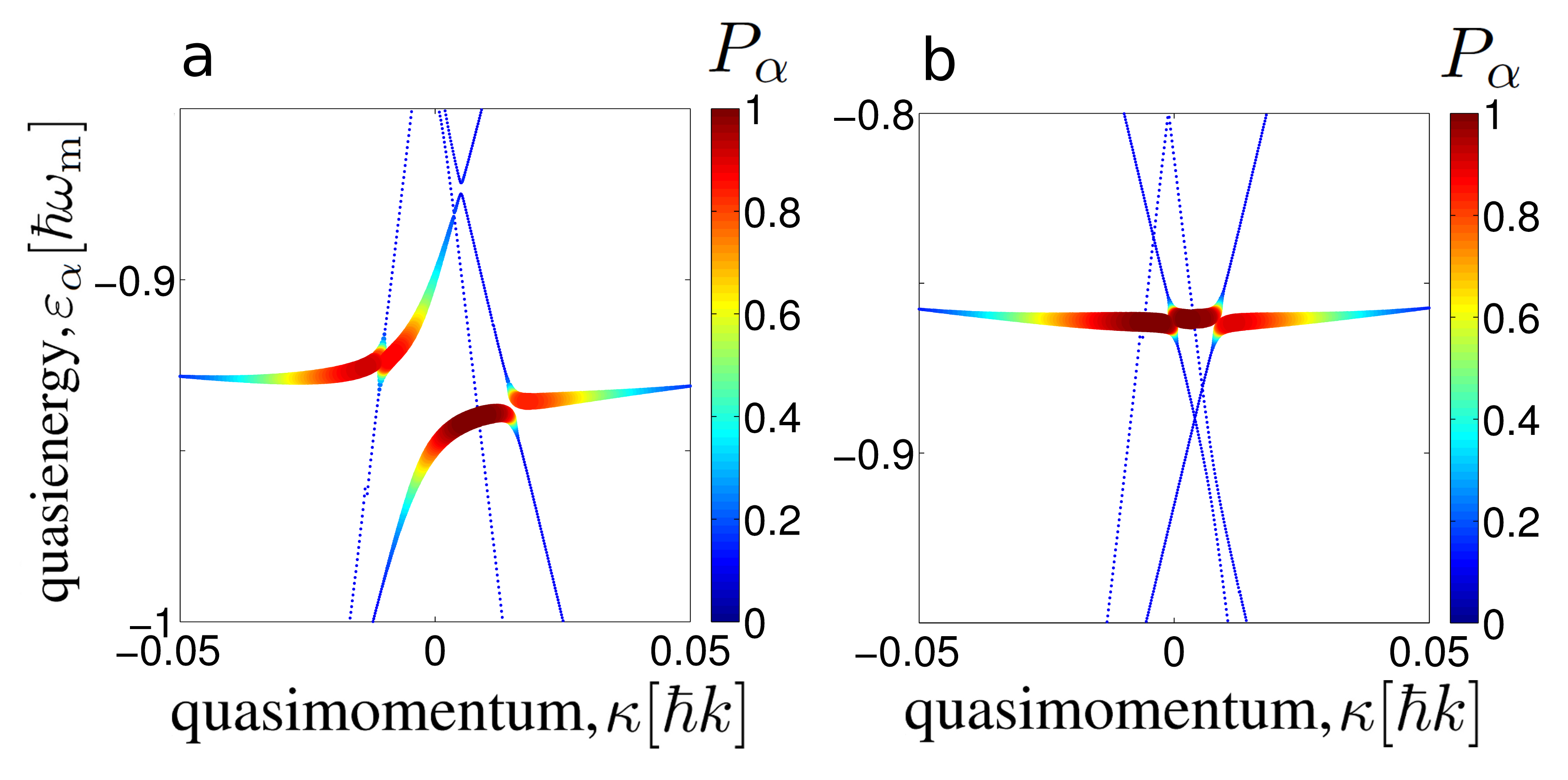}
%\caption{Dependence of computational efficiency measures $R_{100}$ (blue) and $P_{100}$ (red) on the dimension of the problem. Multi-thread version on a single cluster node ($16$ cores with shared memory) is employed (see Table~\ref{table:2}). Here $N_{\rm steps} = 10^2$,  $M = 50$.}
\label{figure:1}
\end{center}
\end{figure}

\textbf{Figure 4 | Population of the Floquet bands}\\
The populations of the Floquet bands of the system~(\ref{Eq:ham}-\ref{Eq:force})
for two values of the modulation amplitude $\Omega_0$ ($\Omega_0=70$ kHz for panel (a) and $\Omega_0=122$ kHz for panel (b)) for a chosen relative phase
$\theta = \pi/2$. Width and color of a band lines encode the relative population
of the Floquet ground band ($\Box$) and ballistic bands ($\triangle$)  by the
initial Gaussian wave packet. The color coding indicating the relative population of the $\alpha$th state (labeled as $P_\alpha$) is given by the intensity bar. 
The remaining parameters used are $V_0 = 3.55 E_{\tn{r}}$, $\beta = 13/7\approx1.86$ and
$\omega_{\tn{m}}=16\omega_{\tn{r}}$.
\newline

\end{document}